\documentclass[a4paper,twosides]{article}
\usepackage{xeCJK,multicol,multirow,graphics,fancyhdr,epstopdf}
\usepackage{amsmath,amsfonts,amssymb,bm,upgreek,mathrsfs,mathcomp}

\usepackage{graphicx}
\usepackage{multicol}
\usepackage{float}
\usepackage{booktabs}
\usepackage{overpic}
\usepackage{orcidlink}
\usepackage{caption}
\usepackage{setspace}

\usepackage[pagewise,switch,columnwise]{lineno}
\usepackage[compress,nospace]{cite}
\usepackage[dvipsnames]{xcolor}
\usepackage{CPL-2023}
\newcommand{\cplyear}{2026} \newcommand{\cplvol}{43}
\newcommand{\cplno}{5} \newcommand{\cplpagenumber}{050201}
 \newcommand{\cplpage}{\cplpagenumber-\thepage}

\begin{document}

\vspace* {-4mm} \begin{center}
\large\bf{\boldmath{Experimental Advances on Light Baryon Spectroscopy at BESIII Experiment}}
\footnotetext{\hspace*{-5.4mm}$^{*}$Corresponding author. Email: wangxiongfei@lzu.edu.cn

\noindent\copyright\,{\cplyear}
\href{http://www.cps-net.org.cn}{Chinese Physical Society} and
\href{http://www.iop.org}{IOP Publishing Ltd}}
\\[5mm]
\normalsize \rm{}Shi Wang(王石)$^{1,2}$~\orcidlink{0000-0003-4624-0117}, Hao Liu(刘昊)$^{1,2}$~\orcidlink{0000-0003-0271-2311}, Shuangshi Fang(房双世)$^{4,5,6}$~\orcidlink{0000-0001-5731-4113}, \\and Xiongfei Wang(王雄飞)$^{1,2,3*}$~\orcidlink{0000-0001-8612-8045}
\\[3mm]\small\sl 
$^1$School of Physical Science and Technology, Lanzhou University, Lanzhou 730000, China

$^2$Lanzhou Center for Theoretical Physics, Key Laboratory of Theoretical Physics of Gansu Province, \\Key Laboratory of Quantum Theory and Applications of MoE, \\Gansu Provincial Research Center for Basic Disciplines of Quantum Physics, \\Lanzhou University, Lanzhou 730000, China

$^{3}$Frontiers Science Center for Rare Isotopes, Lanzhou University, Lanzhou 730000, China

$^{4}$ Institute of High Energy Physics, Beijing 100049, People's Republic of China

$^{5}$ University of Chinese Academy of Sciences, Beijing 100049, China

$^{6}$ Institute of Physics, Henan Academy of Sciences, Zhengzhou 450046, China
\\[4mm]\normalsize\rm{}(Received xxx; accepted manuscript online xxx)
\end{center}
\vskip 1.5mm

\small{\narrower The BESIII experiment is currently the world's only electron-positron collider operating in the tau-charm physical energy region. Since starting data taking in 2009, BESIII has accumulated the world's largest data set in the center-of-mass energy range of 1.84-4.95 GeV, including approximately 10 billion $J/\psi$ events and 3 billion $\psi(3686)$ events, together with extensive data on open-charm hadron pair production near threshold regions. These unique datasets, characterized by high statistics and low background, provide unprecedented experimental conditions for studying light baryon spectroscopy. This article systematically reviews the progress made by BESIII in baryon spectroscopy, with a focus on recent breakthrough achievements, including the discovery of excited nucleon states, $\Lambda$ hyperon states, $\Sigma$ hyperon states, $\Xi$ hyperon states and $\Omega^-$ hyperon states. These results expand the spectrum of baryon excited states and provide crucial experimental support for understanding non-perturbative QCD and resolving the ``missing baryon resonances'' problem.
\par}\vskip 3mm
\normalsize\noindent{\narrower{DOI: \href{http://dx.doi.org/10.1088/0256-307X/\cplvol/\cplno/\cplpagenumber}{10.1088/0256-307X/\cplvol/\cplno/\cplpagenumber}}

\par}\vskip 5mm
\begin{multicols}{2}
\indent \emph{1.~Introduction}. Baryons are hadrons composed of three quarks, and their spectral structure serves as an ideal probe for testing non-perturbative effects in Quantum Chromodynamics (QCD). According to the quark model~\cite{Gell-Mann:1964ewy, Glozman:1995fu,Glozman:1997ag,Capstick:2000qj,Wu:2010jy,FolcoCapossoli:2019imm,Menapara:2021dzi,Guo:2024nrf,wang:2024jyk,Wang:2025dur}, baryons should exhibit a rich spectrum of excited states, analogous to the energy level structure in atoms. However, the number of excited states predicted by theory far exceeds the number observed experimentally, making the ``missing baryon resonances'' problem one of the central puzzles in baryon physics~\cite{Capstick:1992th,Isgur:1978wd,Chen:2008aw}. This discrepancy between theory and experiment suggests that our understanding of non-perturbative QCD remains incomplete, potentially involving complex dynamics not fully captured by the quark model, such as pentaquark components, exotic state contributions, or multi-quark correlation effects.

The BESIII detector at the BEPCII collider, with its low-background environment, high statistics, and broad energy coverage at center-of-mass (c.m.) energy ($\sqrt{s}$) range of $1.84$-$4.95~\mathrm{GeV}$, provides a unique experimental platform for baryon spectroscopy~\cite{BESIII:2020nme}. The advantages of the BESIII experiment are threefold: First, operating as a tau-charm factory, it can collect a large number of charmonium decay events, serving as a clean source for the production of baryon excited states. Second, it can take data near the production thresholds of open-charm hadron pairs, effectively suppressing the background and enabling the full reconstruction of charmed baryons. Third, the detector has excellent charged and neutral particle identification capabilities, combined with advanced reconstruction algorithms, ensuring precise measurements of complex final states. As the only currently operating tau-charm factory, the BESIII experiment not only enables systematic studies of light-flavor baryons including strange baryons, but also provides clean data samples near the production thresholds of charmed baryon pairs, opening new avenues for charmed baryon research. Since data collection began in 2009, the BESIII experiment has achieved a series of breakthroughs in baryon spectroscopy~\cite{BESIII:2012ghz,BESIII:2016ssr,BESIII:2016nix,BESIII:2019dve,BESIII:2021gca}, profoundly advancing our understanding of hadron structure and strong interaction dynamics.

In this article,  we provide a concise overview of current experimental progress in light baryon spectroscopy with a focus on results from BESIII experiment, which includes the excited states of nucleon, $\Lambda$, $\Sigma$, $\Xi$ and $\Omega$. Notably, as BESIII continues to accumulate high-precision data, it has emerged as a leading facility for probing hadron structure and testing non-perturbative nature of strong interaction in light baryon spectroscopy. We therefore highlight its recent advances in this field.

\indent \emph{2.~BESIII Experiment}. Over the past decades, the understanding of excited-state hyperons composed of three light quarks has primarily relied on the analysis of $\bar{K}N$ scattering. Due to the use of outdated detectors with poor resolution, some resonances listed by the PDG~\cite{ParticleDataGroup:2024cfk} have not been further investigated for many years. In recent years, with the operation of the BESIII experiment, the charmonium decays produced in $e^+e^-$ annihilation processes provide unique opportunities for studying baryon resonances. Their low signal-to-noise ratio and background levels, combined with high energy resolution and large data samples, create favorable conditions for in-depth exploration of potential decay channels.

\begin{figure}[H]
    \centering
    \includegraphics[width=0.98\columnwidth]{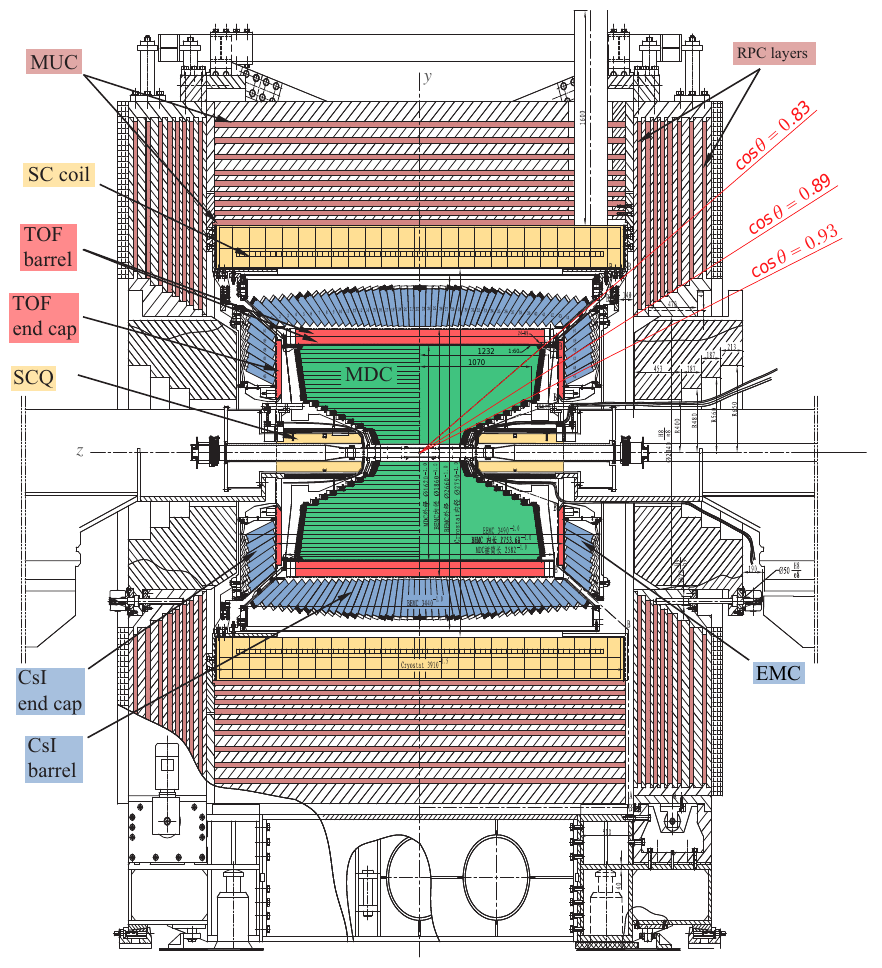}
    \caption{Schematic diagram of the BESIII detector.}
    \label{fig:besiii}
\end{figure}

As schematically illustrated in Fig.~\ref{fig:besiii}, the BESIII detector~\cite{BESIII:2009fln} is designed to record symmetric $e^+e^-$ collisions produced by the BEPCII storage ring~\cite{Yu:2016cof}. Operating at $\tau$-charm physical energy region, i.e. $\sqrt{s} = 1.84$ and $4.95~\mathrm{GeV}$. BESIII so far has acquired large data samples across this energy range~\cite{BESIII:2020nme,Lu:2020imt,Zhang:2022bdc}. The cylindrical core of the BESIII detector encompasses $93\%$ of the full solid angle and comprises a helium-based multilayer drift chamber, a time-of-flight system, and a CsI(Tl) electromagnetic calorimeter. These components are enclosed within a superconducting solenoid magnet that delivers a 1.0 T magnetic field (0.9 T during 2012). An octagonal flux-return yoke, instrumented with resistive plate counters for muon identification interleaved with steel 
\noindent plates, supports the magnet system. The detector provides a charged-particle and photon acceptances of $93\%$ over the full $4\pi$ solid angle. The momentum resolution for charged particles is $0.5\%$ at $1~\mathrm{GeV}/c$, and the $\mathrm{d}E/\mathrm{d}x$ resolution for electrons from Bhabha scattering is $6\%$. In the electromagnetic calorimeter, the energy resolution for photons is $2.5\%$ in the barrel and $5\%$ in the end cap regions at $1~\mathrm{GeV}$. The time resolution of the plastic scintillator TOF system is 68 ps in the barrel and 110 ps in the end caps. The TOF end cap system was upgraded in 2015 using multi-gap resistive plate chamber technology, providing a time resolution of 60 ps, which benefits the data quality~\cite{Li:2017jpg,Guo:2017sjt,Cao:2020ibk}.


\indent \emph{3.~Methodology}. In investigations of excited baryons, one significant challenge arises from their broad widths due to short lifetimes, while another major difficulty is the substantial overlap among different excited states resulting from closely spaced masses. The partial wave analysis (PWA) technique can help address these challenges by disentangling various resonances and determining their properties, including mass, width, spin--parity quantum numbers $J^P$, and partial decay widths. Compared with traditional approaches, the PWA method fully utilizes the four-momentum information of final-state particles. In the PWA, the four-momentum-related amplitude construction will be separated into the angle-dependent part and the energy-dependent part. The angular part can be represented as a rotation matrix in the helicity formalism or included in the tensor form amplitude in the covariant tensor formalism so that the information of $J^P$ of each resonance can be extracted. The energy part is often parameterized using the Breit-Wigner propagator with Blatt-Weisskopf barrier factor~\cite{Chung:1993da}. The total amplitude is then fitted to experimental data to determine parameters such as the magnitude and phase of each partial wave, as well as the mass and width of each resonance. Thus, a cascade amplitude expansion can be applied to any decay process to analyze possible resonances in detail. To construct the full decay amplitude, the helicity formalism is used in conjunction with the isobar model, where the three-body decay is described as a two-step sequential quasi-two-body decay. For each two-body decay $0\to1+2$, the helicity amplitude can be written as
\begin{align}
    A^{0 \to 1+2}_{\lambda_{0},\lambda_{1},\lambda_{2}} = H_{\lambda_{1},\lambda_{2}}^{0 \rightarrow 1+2} D^{J_{0^*}}_{\lambda_{0},\lambda_{1}-\lambda_{2}}(\phi,\theta,0),
\end{align}
where the amplitude $H_{\lambda_{1},\lambda_{2}}^{0\to1+2}$ is given by the $LS$ coupling formula~\cite{Chung:1997jn} along with the barrier factor terms
\begin{align}
    H_{\lambda_{1},\lambda_{2}}^{0 \rightarrow 1+2} =
    \sum_{ls} g_{ls} \sqrt{\frac{2l+1}{2 J_{0}+1}} C^{J_0}_{0,\delta} C^{s}_{\lambda_1,-\lambda_2} q^{l} B_{l}'(q, q_0, d),
\end{align}
where $q$ denotes the breakup momentum in the rest frame of particle 0, $l$ is the orbital angular momentum, $g_{ls}$ are the fit parameters, $J_{0,1,2}$ are the spins of particles 0, 1, and 2, $\lambda_{1,2}$ are the helicities for particles 1 and 2, and $\delta=\lambda_1-\lambda_2$ is the helicity difference. The $C^{J_0}_{0,\delta}=\langle l 0; s \delta|J_{0} \delta\rangle$ and $C^{s}_{\lambda_1,-\lambda_2}=\langle J_{1} \lambda_{1} ;J_{2} (-\lambda_{2}) | s \delta \rangle$ are Clebsch-Gordan Coefficients. The factor $B_{l}'(q, q_0, d)$ is the reduced Blatt-Weisskopf barrier factor~\cite{Chung:1993da}. The amplitude for a complete decay chain is constructed as the product of each two body decay amplitude and the resonant propagator $R$. For example, in the sequential decays $0\to R_{12}+3,\;R_{12}\to1+2$, the amplitude is written as
\begin{align}
    A^{0\to R_{12}+3,\;R_{12}\to1+2}_{\lambda_{0},\lambda_{1},\lambda_{2},\lambda_{3}}
    = \sum_{\lambda_{R_{12}}}A^{0\to R_{12}+3}_{\lambda_{0},\lambda_{R_{12}},\lambda_{3}} R(m) A^{R_{12}\to1+2}_{\lambda_{R_{12}},\lambda_{1},\lambda_{2}}.
\end{align}

\indent \emph{4.~Advances on Light Baryon Spectroscopy}. To gain a deeper understanding of hadron structure and the dynamics of strong interactions, high-precision experimental studies in light baryon spectroscopy are essential. Recent progress, largely driven by the BESIII experiment and other related facilities, has led to significant achievements in this area. These findings provide groundbreaking insights into hadron structure and strong interactions, while also advancing our comprehension of non-perturbative QCD. In the following sections, we will discuss these experimental advances in detail, with charge-conjugate modes implied throughout unless otherwise stated.


\indent \emph{4.1~Nucleon excited states}. In traditional studies of nucleon excited states, using tagged photons or pion beams~\cite{Shklyar:2012js,CBELSATAPS:2015kka,Hunt:2018wqz,CLAS:2018drk}, both the isospin $1/2$ and isospin $3/2$ resonances are excited, further complicating the analysis. An alternative method for studying nucleon resonances involves the decays of charmonium states such as $J/\psi$ and $\psi(3686)$ produced in $e^+e^-$ collider. By focusing on specific decay channels, such as $\psi(3686)\to p\bar{p}\pi^0$, one can study intermediate $N^*$ resonances that couple to $p\pi^0$ or $\bar{p}\pi^0$. In such processes, $\Delta$ resonances are suppressed due to isospin constraints. Moreover, $\Delta$ states typically exhibit significantly broader widths in the mass spectrum, making their separation from background or non-resonant contributions challenging. As a result, $\Delta$ resonances are generally excluded from standard analyses, and the reduced number of states substantially simplifies the procedure~\cite{BES:2001gvq}.

\begin{figure}[H]
    \centering
    \includegraphics[width=1.0\columnwidth]{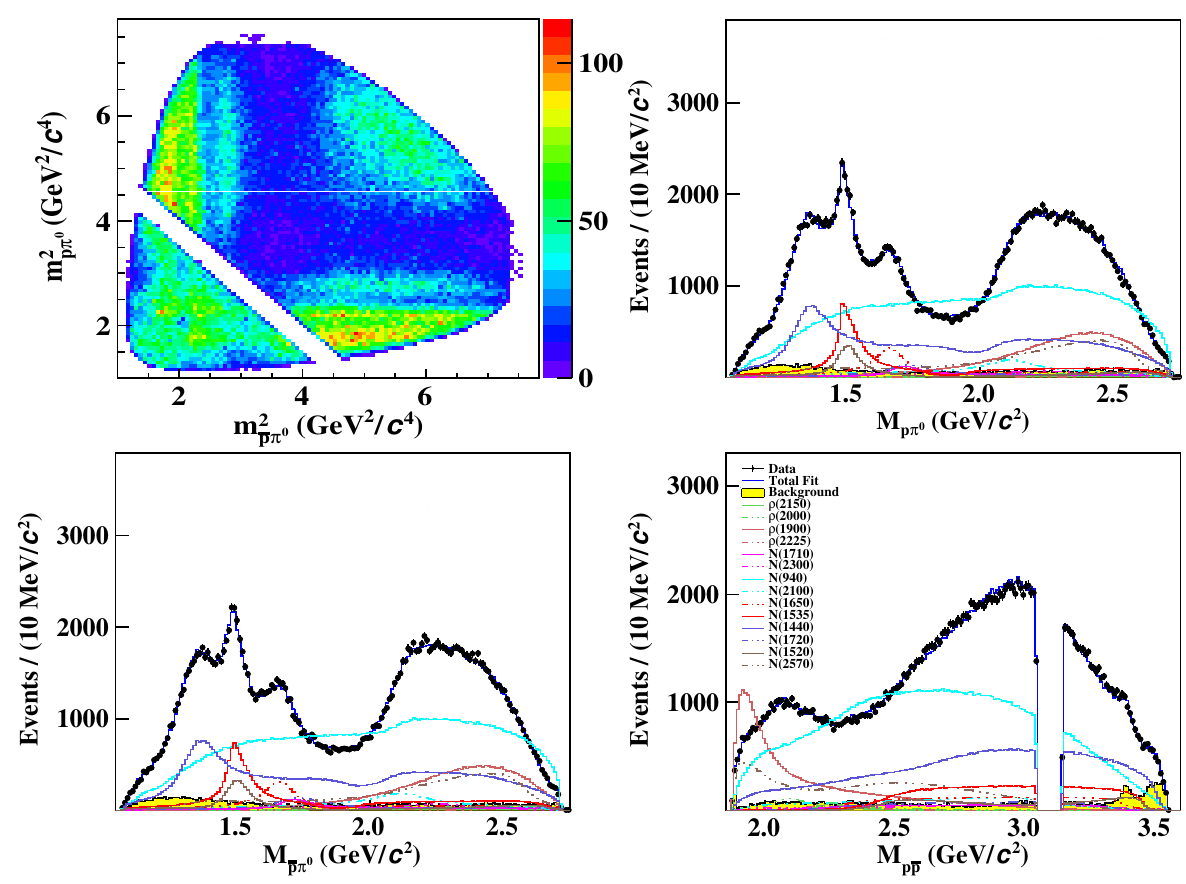}
    \caption{Dalitz plot of $M^2_{p\pi^0}$ versus $M^2_{\bar{p}\pi^0}$ and distributions of $M_{p\bar{p}}$, $M_{p\pi^0}$ and $M_{\bar{p}\pi^0}$ in $\psi(3686)\to p\bar{p}\pi^0$ from BESIII experiment~\cite{BESIII:2024vqu}.}
    \label{fig:pppi0}
\end{figure}

\begin{figure}[H]
    \centering
    \includegraphics[width=1.0\columnwidth]{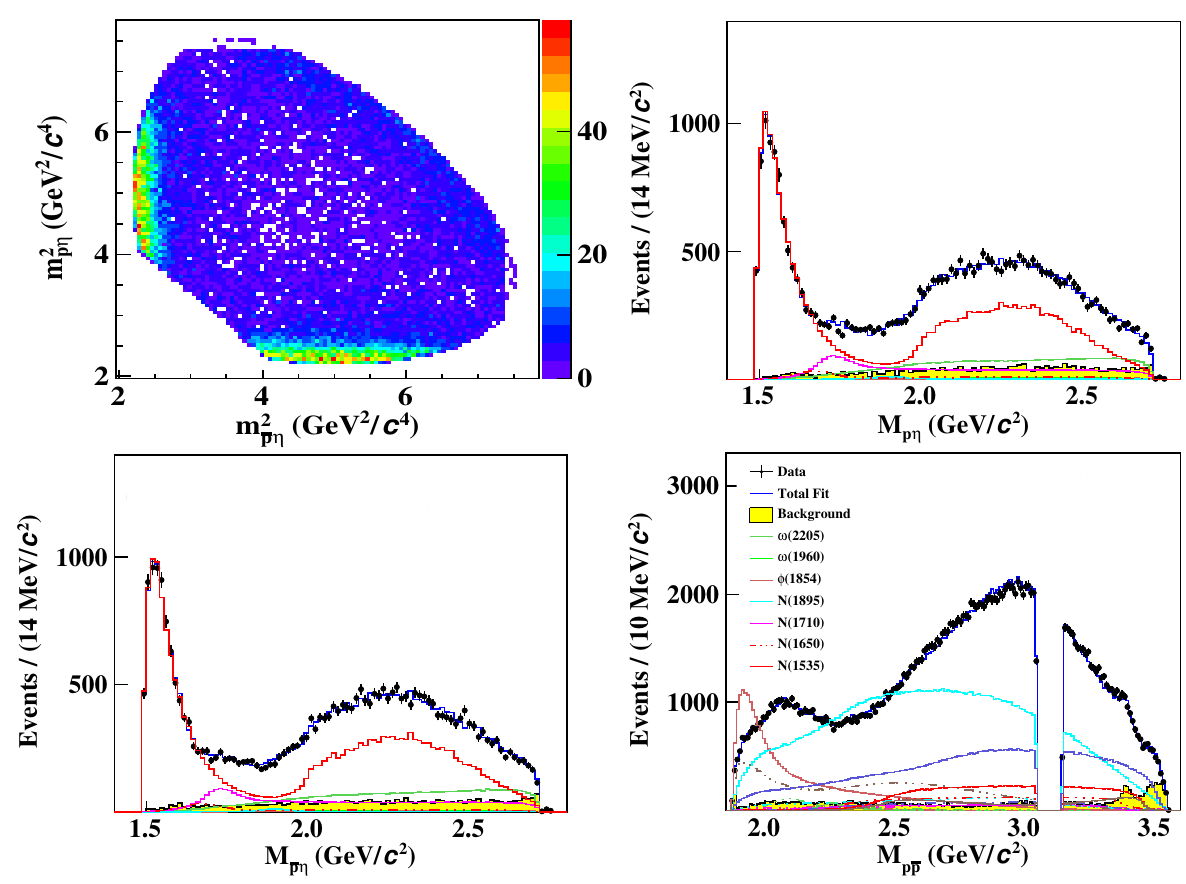}
    \caption{Dalitz plot of $M^2_{p\eta}$ versus $M^2_{\bar{p}\eta}$ and distributions of $M_{p\bar{p}}$, $M_{p\eta}$ and $M_{\bar{p}\eta}$ in $\psi(3686)\to p\bar{p}\eta$ from BESIII experiment~\cite{BESIII:2024vqu}.}
    \label{fig:ppeta}
\end{figure}

The nucleon excited states $N(2300)$ and $N(2570)$, initially observed by BESIII in the decay $\psi(3686)\to p\bar{p}\pi^0$~\cite{BESIII:2012ssm}, have been confirmed in the decay $\psi(3686)\to\bar{p}K^+\Sigma^0$~\cite{BESIII:2026glf}. Their $J^P$ assignments, masses, and widths are all consistent with the PDG~\cite{ParticleDataGroup:2024cfk} and theoretical calculations~\cite{Shah:2018ont}. Recently, BESIII reported the observation of $N^*$ and other excited states via $\psi(3686) \to p \bar{p} \pi^0$ and $p \bar{p} \eta$ using a data sample of 2.7 billion $\psi(3686)$ events~\cite{BESIII:2024vqu}. A PWA was performed in this study, and all resonances were parameterized using the Breit-Wigner function with an energy-dependent width $\Gamma$~\cite{Hunt:2018wqz}. In this analysis, except for several $N^*$ resonances, other states with masses above $1.8~\mathrm{GeV}$ were also determined. No additional states beyond those listed in the PDG~\cite{ParticleDataGroup:2024cfk} were observed. Figures~\ref{fig:pppi0} and~\ref{fig:ppeta} show the Dalitz plots of $M^2_{p\pi^0/\eta}$ versus $M^2_{\bar{p}\pi^0/\eta}$, as well as the invariant mass spectra of $p\pi^0/\eta$ and $\bar{p}\pi^0/\eta$. Table~\ref{tab:n_result} summarizes the results of nucleon excited states from the BESIII experiment.

\begin{table*}
    \centering
    \caption{Summary of mass, width and branching fraction for the nucleon resonances. The first uncertainties are statistical and the second systematic.}
    \label{tab:n_result}
    \begin{tabular}{cc@{\qquad}c@{\qquad}c@{\qquad}c}
        \midrule
        \midrule
        Mode    & Resonance  & Mass $(\mathrm{MeV}/c^2)$ & Width ($\mathrm{MeV}$) & $\mathcal{B}~(\times10^{-6})$ \\
        \midrule
        \multirow{2}{*}{$\psi(3686)\to\bar{p}K^+\Sigma^0$~\cite{BESIII:2026glf}} & $N(2300)$  & $2285.3\pm13.8\pm26.1$ & $335.9\pm14.6\pm21.6$ & $1.95\pm0.53\pm0.58$ \\
            & $N(2570)$  & $2577.9\pm14.8\pm33.2$ & $255.1\pm14.6\pm19.7$ & $2.28\pm0.53\pm0.67$ \\
        \midrule
        \multirow{9}{*}{$\psi(3686)\to p\bar{p}\pi^0$~\cite{BESIII:2024vqu}} & $N(1440)$  & --- & --- & $54.2\pm1.4\pm14.1$ \\
            & $N(1520)$  & --- & --- & $6.6\pm0.4\pm2.0$ \\
            & $N(1535)$  & --- & --- & $17.5\pm0.7\pm3.6$ \\
            & $N(1650)$  & --- & --- & $9.1\pm0.7\pm3.2$ \\
            & $N(1710)$  & --- & --- & $4.2\pm0.5\pm4.0$ \\
            & $N(1720)$  & --- & --- & $6.2\pm0.5\pm2.4$ \\
            & $N(2100)$  & --- & --- & $8.5\pm0.9\pm3.8$ \\
            & $N(2300)$  & --- & --- & $3.6\pm0.6\pm3.0$ \\
            & $N(2570)$  & --- & --- & $19.5\pm0.9\pm14.5$ \\
        \midrule
        \multirow{2}{*}{$\psi(3686)\to p\bar{p}\eta$~\cite{BESIII:2024vqu}} & $N(1535)$  & --- & --- & $50.5\pm1.3\pm7.1$ \\
            & $N(1710)$  & --- & --- & $7.8\pm0.7\pm3.1$ \\
        \midrule
        \midrule
    \end{tabular}
\end{table*}

\indent \emph{4.2~$\Lambda$ excited states}. The $\Lambda$ baryon is an isospin singlet, so the production process is strictly constrained 

\begin{figure}[H]
    \centering
    \includegraphics[width=1.0\columnwidth]{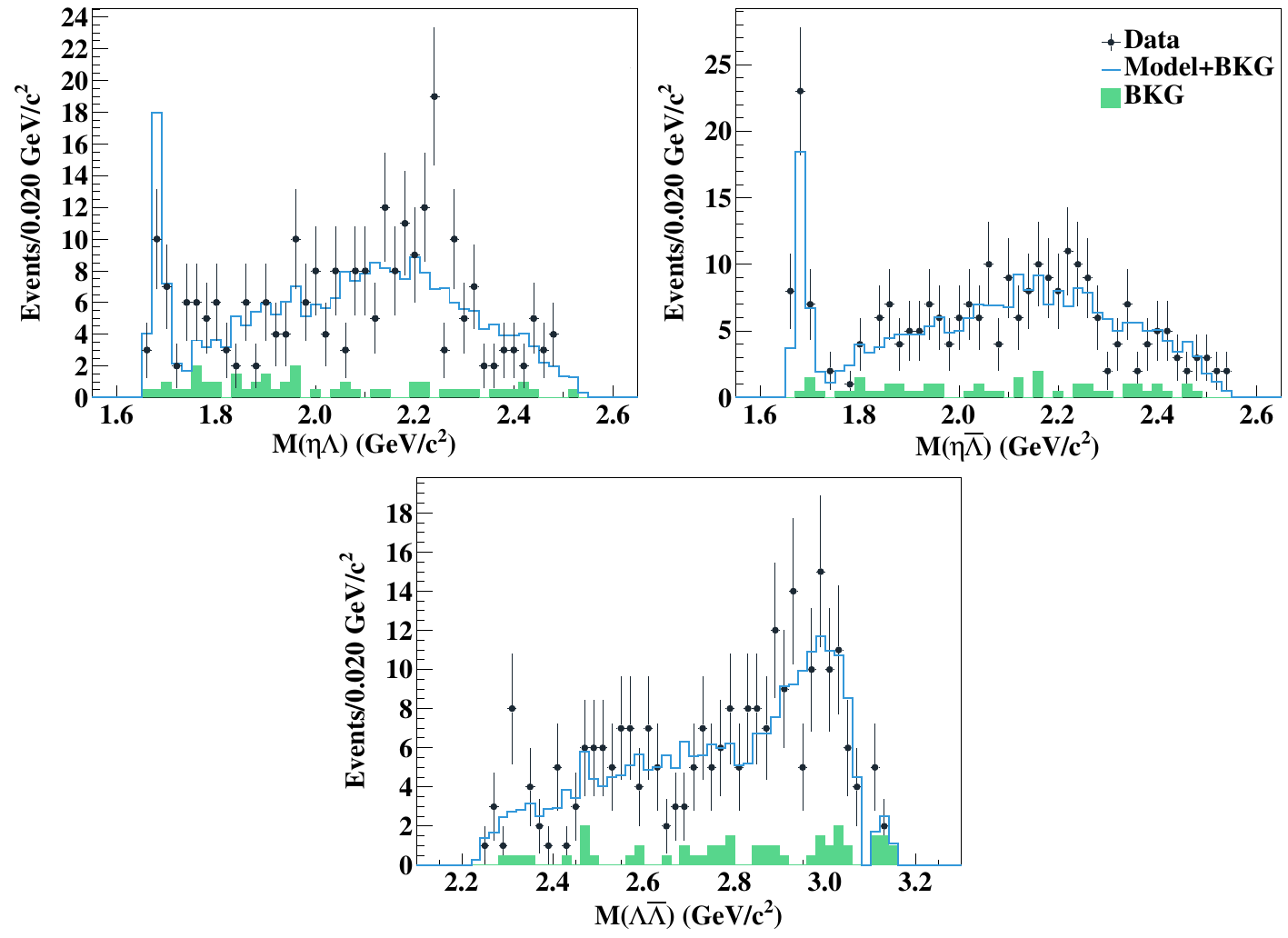}
    \caption{Distributions of $M(\eta\Lambda)$, $M(\eta\bar\Lambda)$ and $M(\Lambda\bar\Lambda)$ in $\psi(3686)\to\Lambda\bar\Lambda\eta$ from BESIII experiment~\cite{BESIII:2022cxi}.}
    \label{fig:etaLambda}
\end{figure}

\begin{figure}[H]
    \centering
    \includegraphics[width=1.0\columnwidth]{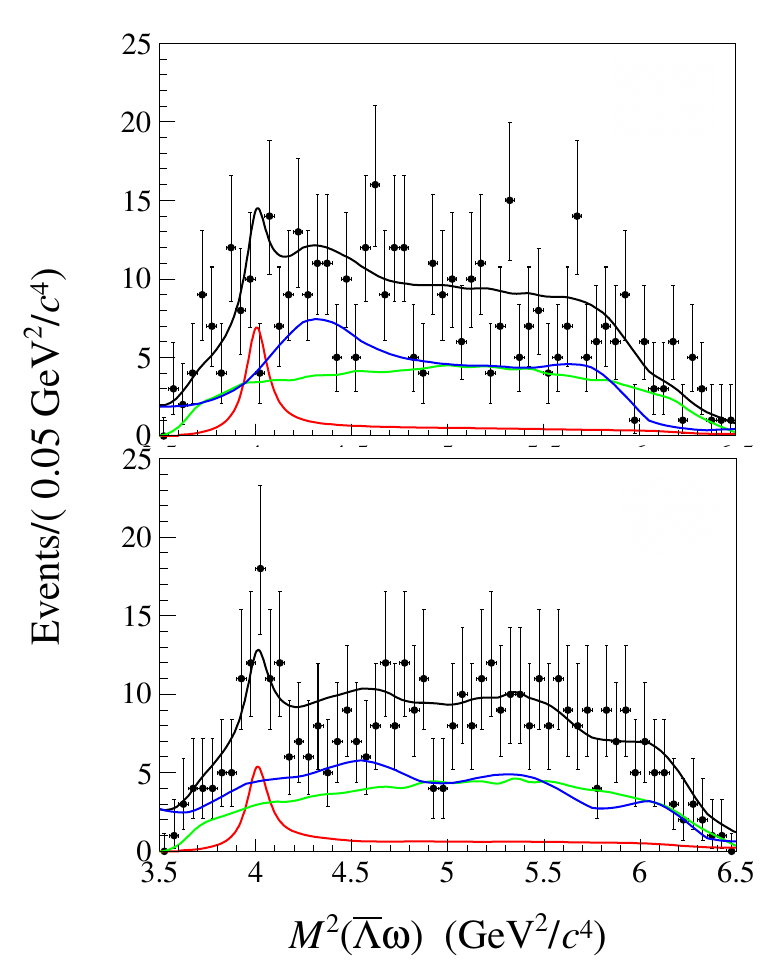}
    \caption{Distributions of $M^2(\Lambda \omega)$ (Top) and $M^2(\bar{\Lambda} \omega)$ (Bottom) in $\psi(3686) \to \Lambda \bar{\Lambda} \omega$ from BESIII experiment~\cite{BESIII:2022fhe}, the red solid curves show the shape of $\Lambda^*$-$\bar{\Lambda}^*$ resonances, the blue solid curves show the background described by $\omega$ sidebands and the green solid curves show the shapes from the non-resonant decay.}
    \label{fig:omegaLambda}
\end{figure}

\begin{figure}[H]
    \centering
    \includegraphics[width=1.0\columnwidth]{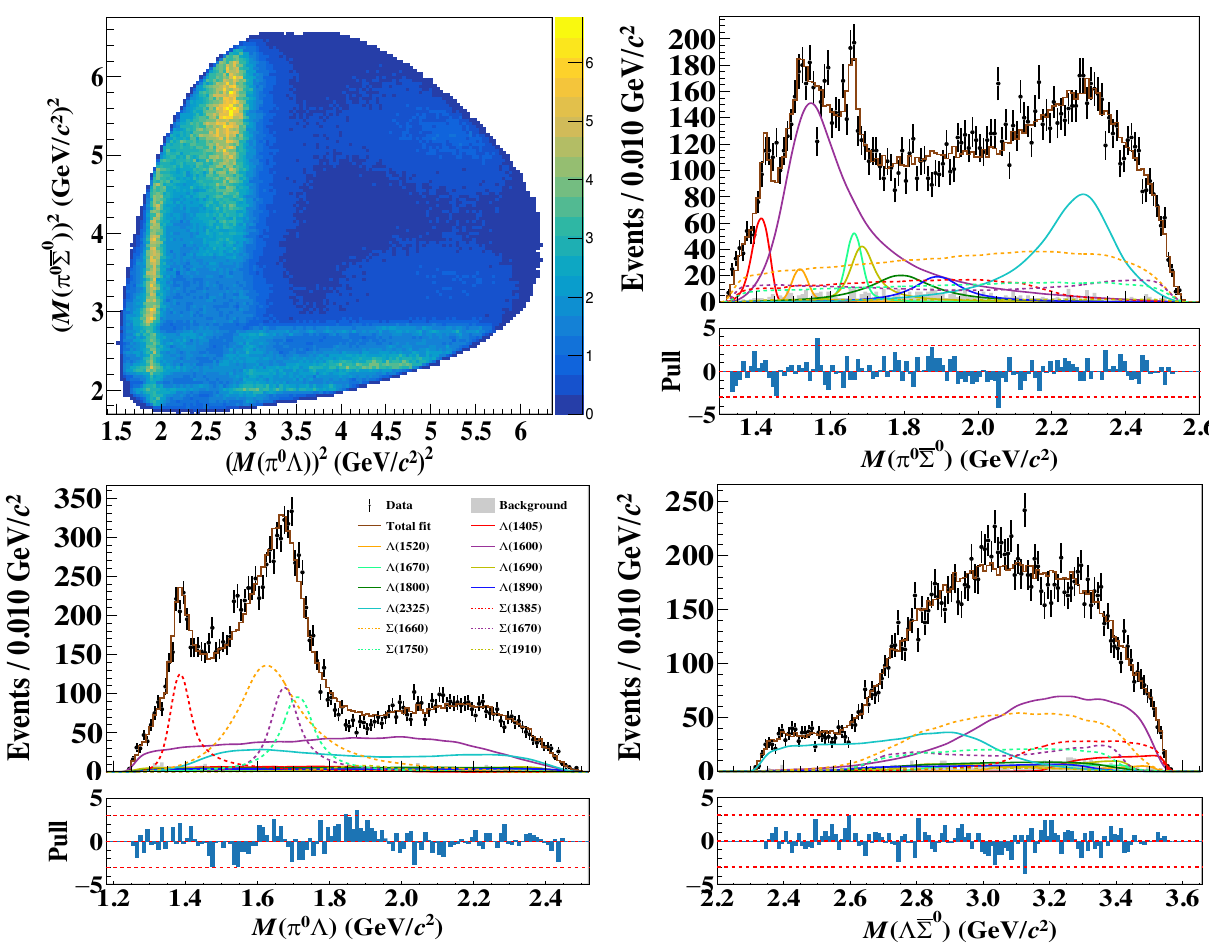}
    \caption{Dalitz plot of $M^2(\pi^0 \Lambda)$ versus $M^2(\pi^0\bar\Sigma^0)$ and distributions of $M(\Lambda\bar\Sigma^0)$, $M(\pi^0 \Lambda)$ and $M(\pi^0\bar\Sigma^0)$ in $\psi(3686)\to\Lambda\bar\Sigma^0\pi^0$ from BESIII experiment~\cite{BESIII:2024jgy}.}
    \label{fig:LSpi0}
\end{figure}

\noindent if isospin symmetry is conserved. Benefiting from the large data sample and suitable phase-space coverage, several resonances have been observed via PWA at BESIII. In 2022, BESIII studied the decay  $\psi(3686) \to \Lambda \bar{\Lambda} \eta$ and identified the $\Lambda(1670)$ resonance near the threshold in the $\eta \Lambda/\bar{\Lambda}$ invariant mass spectrum~\cite{BESIII:2022cxi}. The corresponding distributions are shown in Fig.~\ref{fig:etaLambda}. The measured mass and width are  $(1672 \pm 5 \pm 6)~\mathrm{MeV}/c^2$ and $(38 \pm 10 \pm 19)~\mathrm{MeV}$, respectively, and the spin-parity $J^P$ is determined to be $1/2^-$, consistent with the PDG values~\cite{ParticleDataGroup:2024cfk}. In the same year, a study of $\psi(3686)\to\Lambda\bar\Lambda\omega$~\cite{BESIII:2022fhe} reported a new resonance labeled $\Lambda(2000)$, as shown in Fig.~\ref{fig:omegaLambda}, with a significance of only $3\sigma$. Its extracted width is considerably narrower than any known state listed in the PDG~\cite{ParticleDataGroup:2024cfk}, suggesting the need for further investigation. In 2023, studies of  $e^+ e^- \to p K^- \bar{\Lambda}$ at $4.178~\mathrm{GeV}$~\cite{BESIII:2023kgz} and $J/\psi \to \Sigma^{\mp}\bar\Lambda\pi^\pm$~\cite{BESIII:2023syz} did not reveal or measure any $\Lambda^*$ resonances, though the former analysis reported a new $1^+$ resonance, denoted $K_1(2085)$, in the $p\bar{\Lambda}$ mass spectrum. In 2025, the BESIII collaboration reported a significant signal of the $\Lambda(2325)$ state in the decay $\psi(3686) \to \Lambda \bar{\Sigma}^0 \pi^0$~\cite{BESIII:2024jgy}. The corresponding mass spectrum fit curves are shown in Fig.~\ref{fig:LSpi0}. Additionally, the two-pole structure of $\Lambda(1405)$ was investigated using two independent parameterizations: a Flatt\'e-like model~\cite{Chung:1995dx} and a chiral dynamics approach~\cite{Jido:2003cb}. However, due to limited statistics, it was not possible to determine which model provides a better description of the data. Therefore, both parameterization results are reported in this analysis.

\begin{figure}[H]
    \centering
    \includegraphics[width=0.48\columnwidth]{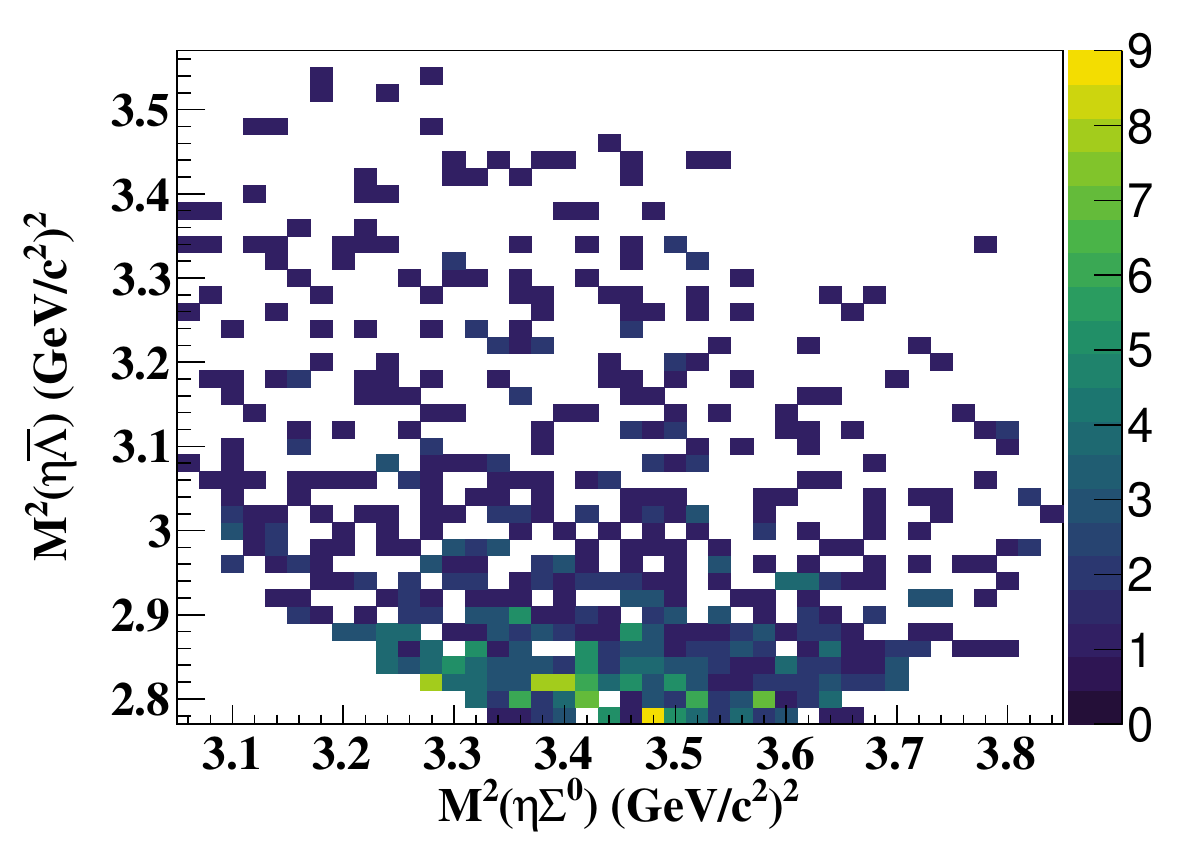}
        \includegraphics[width=0.48\columnwidth]{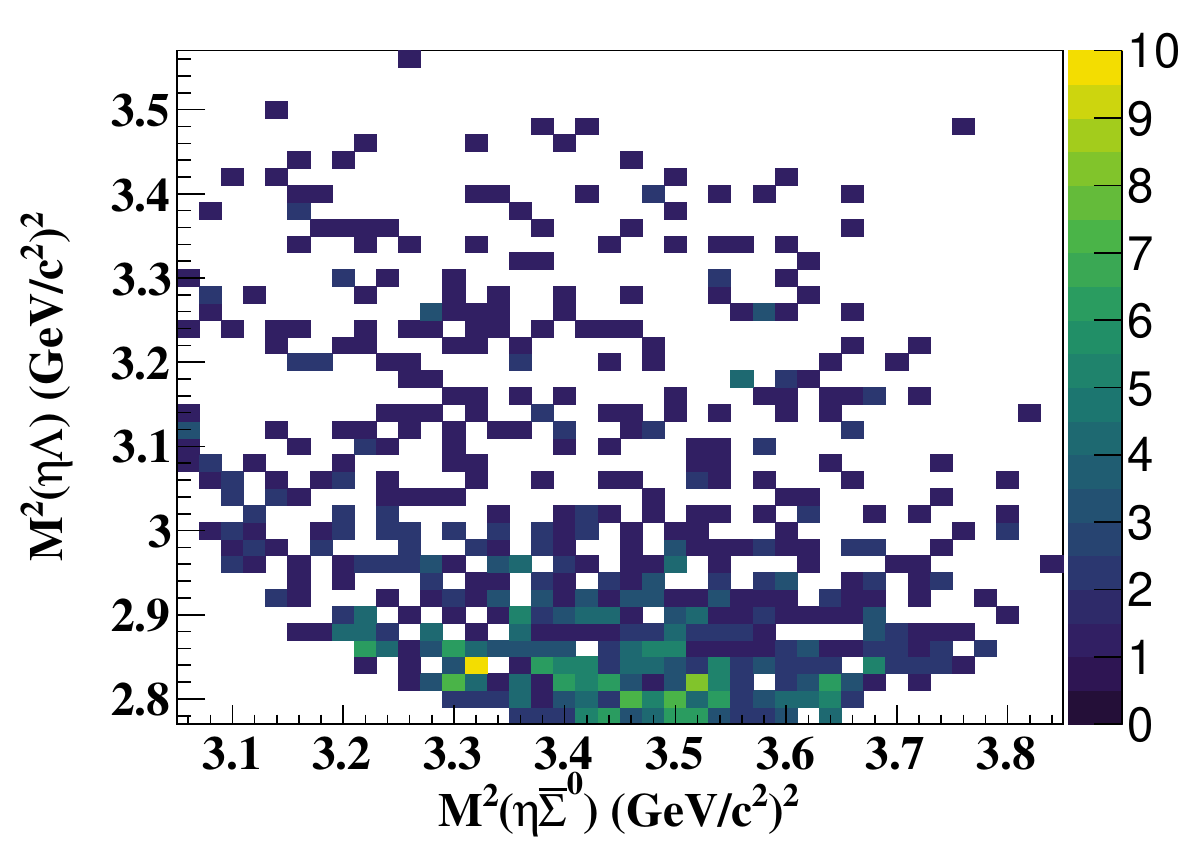}
    \caption{Dalitz plots of $M^2(\eta\Lambda)$ versus $M^2(\eta\bar\Sigma^0)$ in $\psi(3686)\to\Lambda \bar\Sigma^0\eta$ from BESIII experiment~\cite{BESIII:2026vrm}.}
    \label{fig:etaLS}
\end{figure}

Recently, the BESIII collaboration has reported observations of $\Lambda(1690)$ and $\Lambda(1810)$ in the isospin-violating process $J/\psi \to \Lambda\bar{\Sigma}^0\eta$~\cite{BESIII:2026vrm}, as illustrated in the Dalitz plots in Fig.~\ref{fig:etaLS}. The measured mass of $\Lambda(1810)$ is approximately $90~\mathrm{MeV}/c^2$ higher than the PDG-accepted value of $1790~\mathrm{MeV}/c^2$~\cite{ParticleDataGroup:2024cfk}, though the results remain consistent within two standard deviations. This measured mass is also consistent with predictions from quark model calculations~\cite{Loring:2001ky}. Given the proximity of its mass to that of $\Lambda(1890)$ with $J^P=3/2^+$, an alternative fit was performed by replacing $\Lambda(1810)$ with $\Lambda(1890)$, which led to a slightly worse fit quality. Based on the currently limited statistics, we conclude that the presence of $\Lambda(1890)$ cannot be ruled out. Table~\ref{tab:l_result} summarizes the recent results of $\Lambda$ excited states from the BESIII experiment.

\indent \emph{4.3~$\Sigma$ excited states}. Recent advances in our understanding of $\Sigma$ excited states have been driven largely 

\noindent by multi-channel analyses of $\bar{K}N$ scattering data~\cite{Zhang:2013sva,Sarantsev:2019xxm}. Nevertheless, several predicted $\Sigma$ resonances remain unobserved compared to theoretical expectations~\cite{Menapara:2024wpb}. With the world's largest data samples of 

\noindent $J/\psi$ and $\psi(3686)$ events, the BESIII experiment offers a unique opportunity to systematically investigate these missing $\Sigma$ states. 

\begin{table*}
    \centering
    \caption{Summary of mass, width and branching fraction for the $\Lambda$ resonances. The first uncertainties are statistical and the second systematic.}
    \label{tab:l_result}
    \begin{tabular}{cc@{\qquad}c@{\qquad}c@{\qquad}c}
        \midrule
        \midrule
        Mode     & Resonance & Mass $(\mathrm{MeV}/c^2)$ & Width ($\mathrm{MeV}$) & $\mathcal{B}~(\times10^{-5})$ \\
        \midrule
        $\psi(3686)\to\Lambda\bar{\Lambda}\eta$~\cite{BESIII:2022cxi} & $\Lambda(1670)$  & $1672\pm5\pm6$ & $38\pm10\pm19$ & $1.29\pm0.31\pm0.62$ \\
        \midrule
        $\psi(3686)\to\Lambda\bar{\Lambda}\omega$~\cite{BESIII:2022fhe} & $\Lambda^*$  & $2001\pm7$ & $36\pm14$ & $<1.40$ (@ $90\%$ CL) \\
        \midrule
        \multirow{8}{*}{$\psi(3686)\to\Lambda\bar{\Sigma}^0\pi^0$~\cite{BESIII:2024jgy}} & $\Lambda(1405)$ & --- & --- & $0.44\pm0.05\pm0.12$ \\
         & $\Lambda(1520)$ & $1519.9\pm1.6\pm4.8$ & $20.6\pm1.9\pm0.6$ & $0.19\pm0.05\pm0.05$ \\
         & $\Lambda(1600)$ & $1570.5\pm4.6\pm12.1$ & $228.1\pm11.9\pm34.8$ & $4.49\pm0.25\pm1.23$ \\
         & $\Lambda(1670)$ & $1667.5\pm2.3\pm3.6$ & $30.2\pm1.2\pm1.8$ & $0.33\pm0.07\pm0.09$ \\
         & $\Lambda(1690)$ & $1691.1\pm4.4\pm15.7$ & $72.3\pm4.7\pm18.3$ & $0.55\pm0.11\pm0.10$ \\
         & $\Lambda(1800)$ & $1800.9\pm2.3\pm4.2$ & $13.3\pm2.5\pm4.5$ & $12.6\pm2.5\pm4.5$ \\
         & $\Lambda(1890)$ & $1897.2\pm2.3\pm4.2$ & $9.6\pm2.5\pm4.5$ & $6.2\pm2.5\pm4.5$ \\
         & $\Lambda(2325)$ & $2306.5\pm2.3\pm4.2$ & $6.3\pm2.5\pm4.5$ & $17.1\pm2.5\pm4.5$ \\
        \midrule
        \multirow{2}{*}{$\psi(3686)\to\Lambda\bar{\Sigma}^0\eta$~\cite{BESIII:2026vrm}} & $\Lambda(1670)$  & $1668.8\pm1.2\pm1.5$ & $3.1\pm0.2\pm0.3$ & $21.2\pm0.8\pm1.4$ \\
         & $\Lambda(1810)$  & $1881.5\pm2.3\pm4.2$ & $16.5\pm0.7\pm0.9$ & $20.3\pm0.7\pm1.4$ \\
        \midrule
        $\psi(3686)\to\bar{p}K^+\Sigma^0$~\cite{BESIII:2026glf} & $\Lambda(1520)$ & --- & --- & $0.49\pm0.05\pm0.14$ \\
        \midrule
        \midrule
    \end{tabular}
\end{table*}

\begin{figure}[H]
    \centering
    \begin{overpic}[width=1.0\columnwidth]{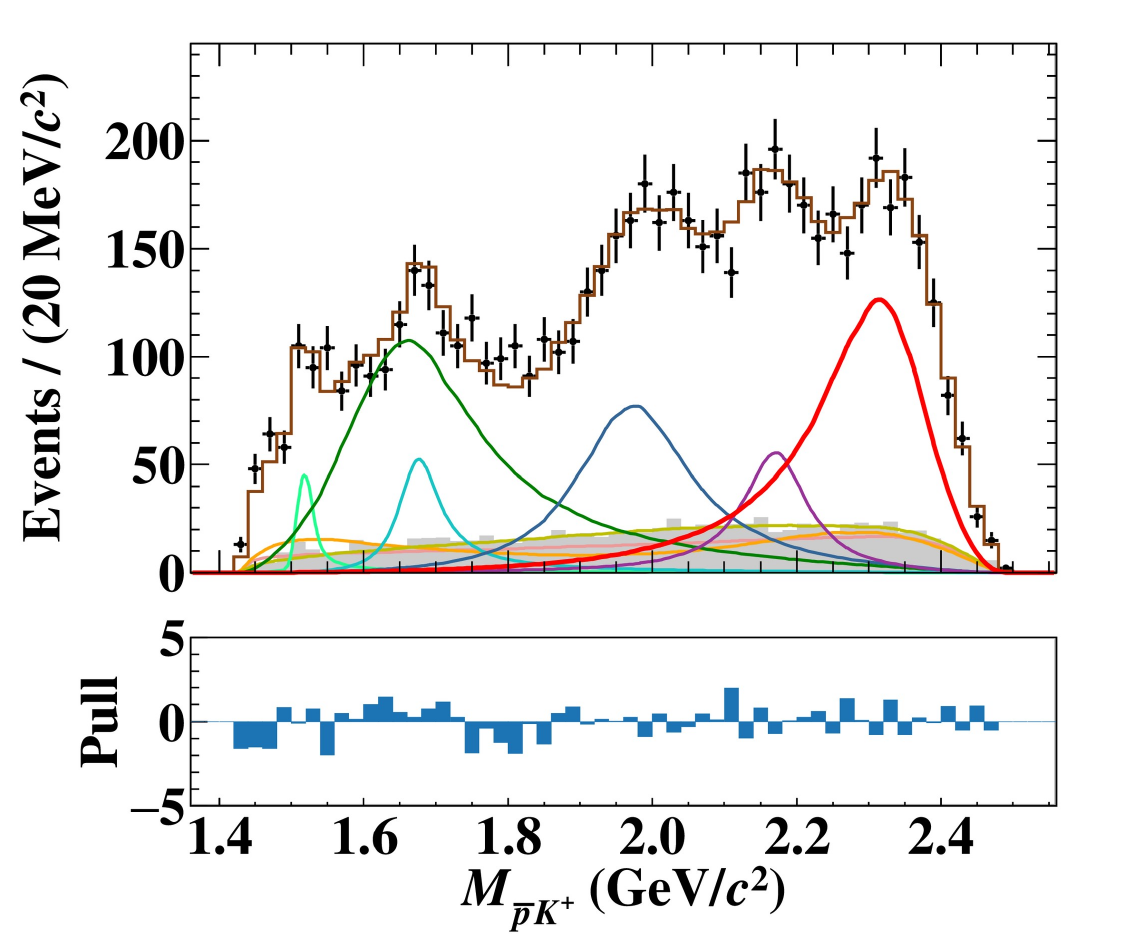}
        \put(20,65){\includegraphics[clip,trim=2cm 7cm 2.5cm 1.5cm,width=0.3\columnwidth]{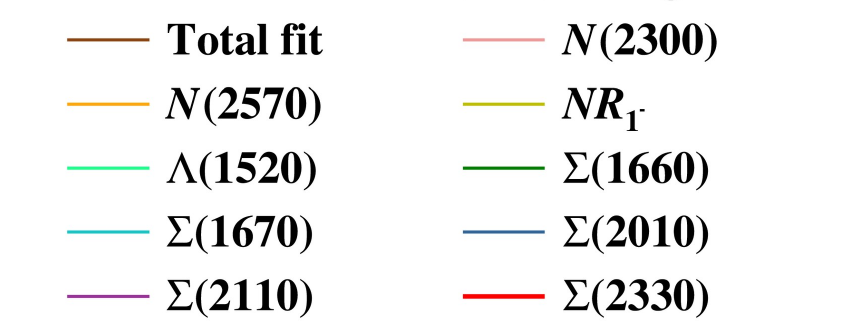}}
    \end{overpic}
    \caption{Distribution of $M_{\bar{p}K^+}$ from BESIII experiment~\cite{BESIII:2026glf}.}
    \label{fig:pksp_m}
\end{figure}

\indent In the study of $\Sigma$ baryons, the BESIII collaboration analyzed the decay $\psi(3686)\to\bar{p}K^+\Sigma^0$~\cite{BESIII:2026glf}, and identified a new excited $\Sigma$ state, $\Sigma(2330)$, with a statistical significance of $11.9\sigma$ in the $M_{\bar{p}K^+}$ distribution, as illustrated in Fig.~\ref{fig:pksp_m}. The measured mass and width are $(2334.7\pm7.9\pm16.0)~\mathrm{MeV}/c^2$ and $(206.3\pm9.5\pm18.4)~\mathrm{MeV}$, respectively, where the first uncertainty is statistical and the second systematic. The spin-parity assignment for $\Sigma(2330)$ favors $3/2^-$. No known resonance in the PDG~\cite{ParticleDataGroup:2024cfk} corresponds to both its mass and width. A recent theoretical prediction~\cite{Menapara:2024wpb} attributes excited $\Sigma$ states near $2.33~\mathrm{GeV}/c^2$ to the 1F family. Our result is consistent with the predicted 1F$(3/2^-)$ state within $5~\mathrm{MeV}/c^2$. Alternative spin-parity assignments of $3/2^+$ and $5/2^-$ yield significances of $10.7\sigma$ and $11.0\sigma$, respectively-values very close to that of the nominal $3/2^-$ hypothesis. Due to limited statistics and the proximity of the $\Sigma(2330)$ mass to the phase-space threshold, the current data cannot definitively discriminate among these possibilities. Further studies with larger datasets and additional decay channels will be necessary to unam-

\noindent biguously determine the quantum numbers of $\Sigma(2330)$. Furthermore, spin-parity analyses were carried out for the one-star states $\Sigma(2010)$ and $\Sigma(2110)$. The results support $J^P=3/2^-$ and $J^P=1/2^-$, respectively, in agreement with the assignments given by the PDG~\cite{ParticleDataGroup:2024cfk}.

\begin{figure}[H]
    \centering
    \includegraphics[width=1.0\columnwidth]{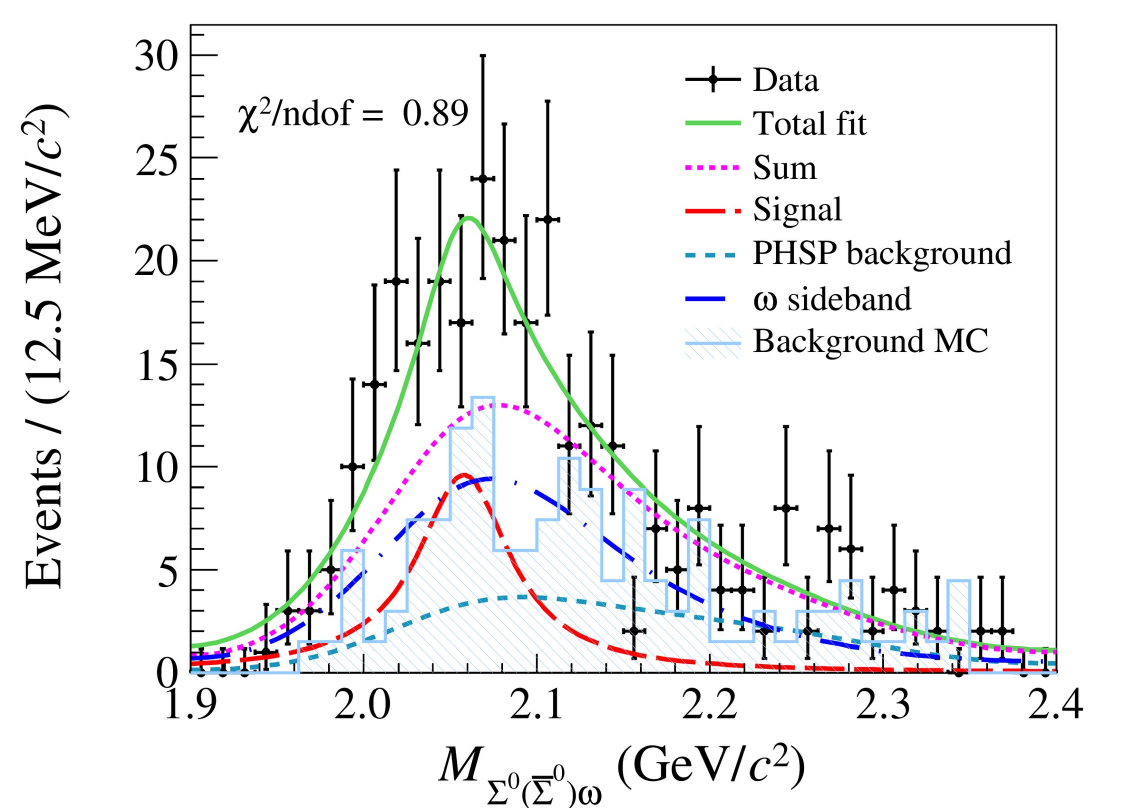}
    \caption{Distribution of $M_{\Sigma^0(\bar{\Sigma}^0)\omega}$ from BESIII experiment~\cite{BESIII:2025gzs}.}
    \label{fig:ssomega_m}
\end{figure}

In the analyses of $J/\psi \to \Sigma^{\mp} \bar{\Lambda} \pi^{\pm}$~\cite{BESIII:2023syz} and $\psi(3686) \to \Lambda \bar{\Sigma}^0 \pi^0$~\cite{BESIII:2024jgy}, several resonances $\Sigma^*$ are considered in the nominal fit according to the PDG~\cite{ParticleDataGroup:2024cfk}, but no new states have been confirmed. While in the analysis $\psi(3686)\to\Sigma^0\bar{\Sigma}^0\omega$~\cite{BESIII:2025gzs}, a hint of a resonance in the $\Sigma^0(\bar{\Sigma}^0)\omega$ invariant mass spectrum is observed, as shown in Fig.~\ref{fig:ssomega_m}. The fit to the invariant mass distribution yields a significance of $2.5\sigma$, with a measured mass and width of $(2058\pm15)~\mathrm{MeV}/c^2$ and $(65\pm18)~\mathrm{MeV}$, respectively. However, due to the limited statistics, the intermediate state is not analyzed through PWA method, and its quantum numbers cannot be determined.

\begin{figure}[H]
    \centering
    \includegraphics[width=1.0\columnwidth]{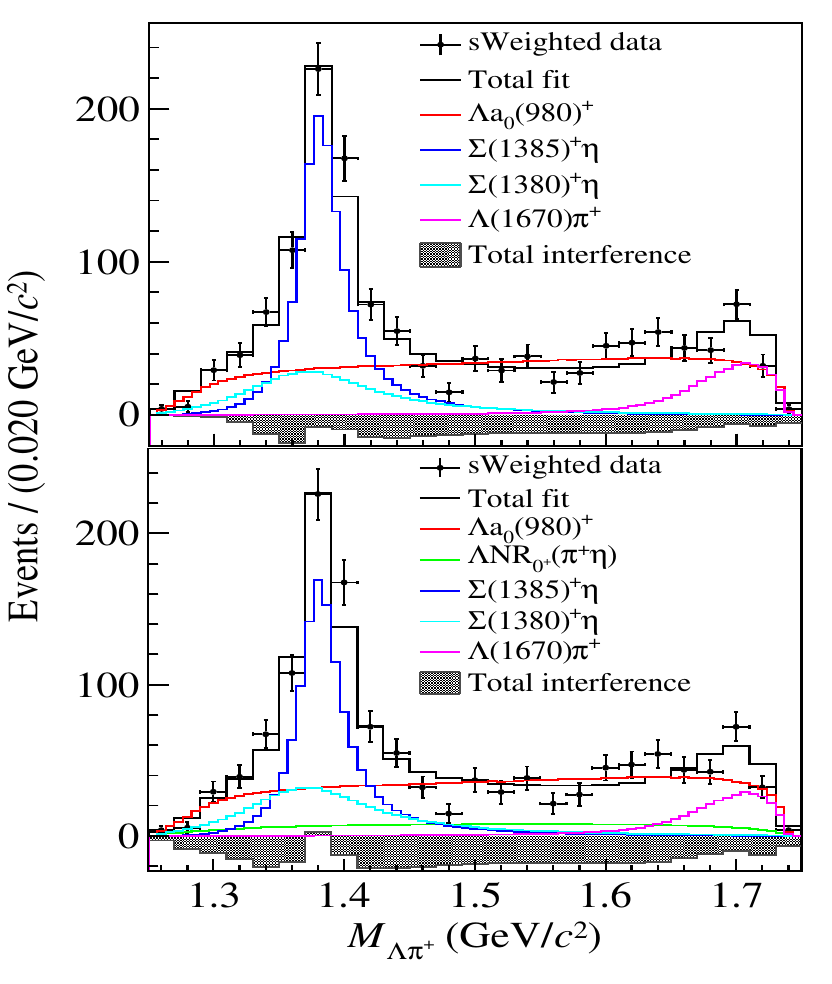}
    \caption{Distribution of $M_{\Lambda\pi^+}$ with PWA fits using model A (Top) and model B (Bottom) from BESIII experiment~\cite{BESIII:2024mbf}.}
    \label{fig:lc_1380_2m}
\end{figure}

In 2025, the BESIII collaboration reported the first evidence for a potential pentaquark state, the $\Sigma(1380)^+$, identified in the $\Lambda\pi^+$ system from the decay $\Lambda_c^+\to\Lambda\pi^+\eta$~\cite{BESIII:2024mbf} using a PWA. In this analysis, the Breit-Wigner mass and width of the $\Sigma(1380)^+$ were fixed at $1380~\mathrm{MeV}/c^2$ and $120~\mathrm{MeV}$, respectively, following predictions from Refs.~\cite{Wu:2009tu,Xie:2017xwx}. To conservatively evaluate the statistical significance of the $\Sigma(1380)^+$ signal, two models were compared: ``model A'', which includes the resonances $\Lambda a_0(980)^+$, $\Sigma(1385)^+\eta$, $\Lambda(1670)\pi^+$, and $\Sigma(1380)^+\eta$; and ``model B'', which adds a non-resonant component with $J^P=0^+$ ($\Lambda\mathrm{NR}_{0^+}$) to the resonance set of model A. The corresponding fit curves are presented in Fig.~\ref{fig:lc_1380_2m}. Based on the likelihood differences between fits with and 

\noindent without the $\Sigma(1380)^+$ resonance, the statistical significances were found to be $6.1\sigma$ for model A and $3.3\sigma$ for model B. Table~\ref{tab:s_result} summarizes the recent results of $\Sigma$ excited states from the BESIII experiment.

\begin{table*}
    \centering
    \caption{Masses, widths, and production branching fractions of the $\Sigma$ resonances. The first uncertainties are statistical and the second systematic.}
    \label{tab:s_result}
    \begin{tabular}{cc@{\qquad}c@{\qquad}c@{\qquad}c}
        \midrule
        \midrule
        Mode & Resonance & Mass $(\mathrm{MeV}/c^2)$ & Width ($\mathrm{MeV}$) & $\mathcal{B}~(\times10^{-5})$ \\
        \midrule
        \multirow{5}{*}{$\psi(3686)\to\bar{p}K^+\Sigma^0$~\cite{BESIII:2026glf}} & $\Sigma(1660)$ & $1680.5\pm13.8\pm21.3$ & $247.7\pm15.8\pm18.8$ & $5.84\pm0.69\pm1.55$ \\
         & $\Sigma(1670)$ & $1680.5\pm4.1\pm11.0$ & $73.4\pm9.1\pm7.1$  & $1.22\pm0.38\pm0.28$ \\
         & $\Sigma(2010)$ & $1980.1\pm10.0\pm18.1$ & $192.5\pm13.3\pm13.3$ & $3.11\pm0.50\pm0.92$ \\
         & $\Sigma(2110)$ & $2172.4\pm6.3\pm10.9$ & $107.2\pm9.3\pm10.2$ & $1.17\pm0.23\pm0.38$ \\
         & $\Sigma(2330)$ & $2334.7\pm7.9\pm16.0$ & $206.3\pm9.5\pm18.4$ & $4.47\pm0.58\pm1.52$ \\
        \midrule
        \multirow{5}{*}{$\psi(3686)\to\Lambda\bar{\Sigma}^0\pi^0$~\cite{BESIII:2024jgy}} & $\Sigma(1385)$ & $1388.2\pm1.9\pm3.3$ & $60.5\pm3.6\pm6.3$  & $1.30\pm0.11\pm0.19$ \\
         & $\Sigma(1660)$ & $1643.2\pm4.5\pm7.6$ & $221.3\pm13.1\pm41.1$ & $3.53\pm0.29\pm0.63$ \\
         & $\Sigma(1670)$ & $1679.7\pm3.4\pm4.3$ & $87.0\pm6.4\pm8.9$ & $1.60\pm0.25\pm0.38$ \\
         & $\Sigma(1750)$ & $1714.9\pm4.2\pm7.3$ & $97.2\pm9.8\pm9.7$ & $1.39\pm0.23\pm0.32$ \\
         & $\Sigma(1910)$ & $1912.1\pm10.6\pm33.6$ & $225.1\pm24.5\pm46.0$ & $0.23\pm0.10\pm0.06$ \\
        \midrule
        $\psi(3686)\to\Sigma^0\bar{\Sigma}^0\omega$~\cite{BESIII:2025gzs}     & $\Sigma^*$  & $2058\pm15$ & $36\pm14$ & --- \\
        \midrule
        \midrule
    \end{tabular}
\end{table*}

\indent \emph{4.4~$\Xi$ excited states}. The $\Xi$ baryon is a doubly strange hyperon composed of two strange quarks and one light quark. Compared to singly strange hyperons ($\Lambda$, $\Sigma$), $\Xi$ baryons have smaller production cross-sections, making their experimental investigation more challenging. For the excited states of the $\Xi$ baryon, experimental evidence remains limited, with the exception of the well-established $\Xi(1530)$. Resonances in the mass region of $1.5$ and $1.61~\mathrm{GeV}$ can be easily identified via the $\Xi \pi$ decay channel, while those with masses above $1.61~\mathrm{GeV}$ may appear in decay 

\begin{figure}[H]
	\begin{center}
	\includegraphics[width=1.0\columnwidth]{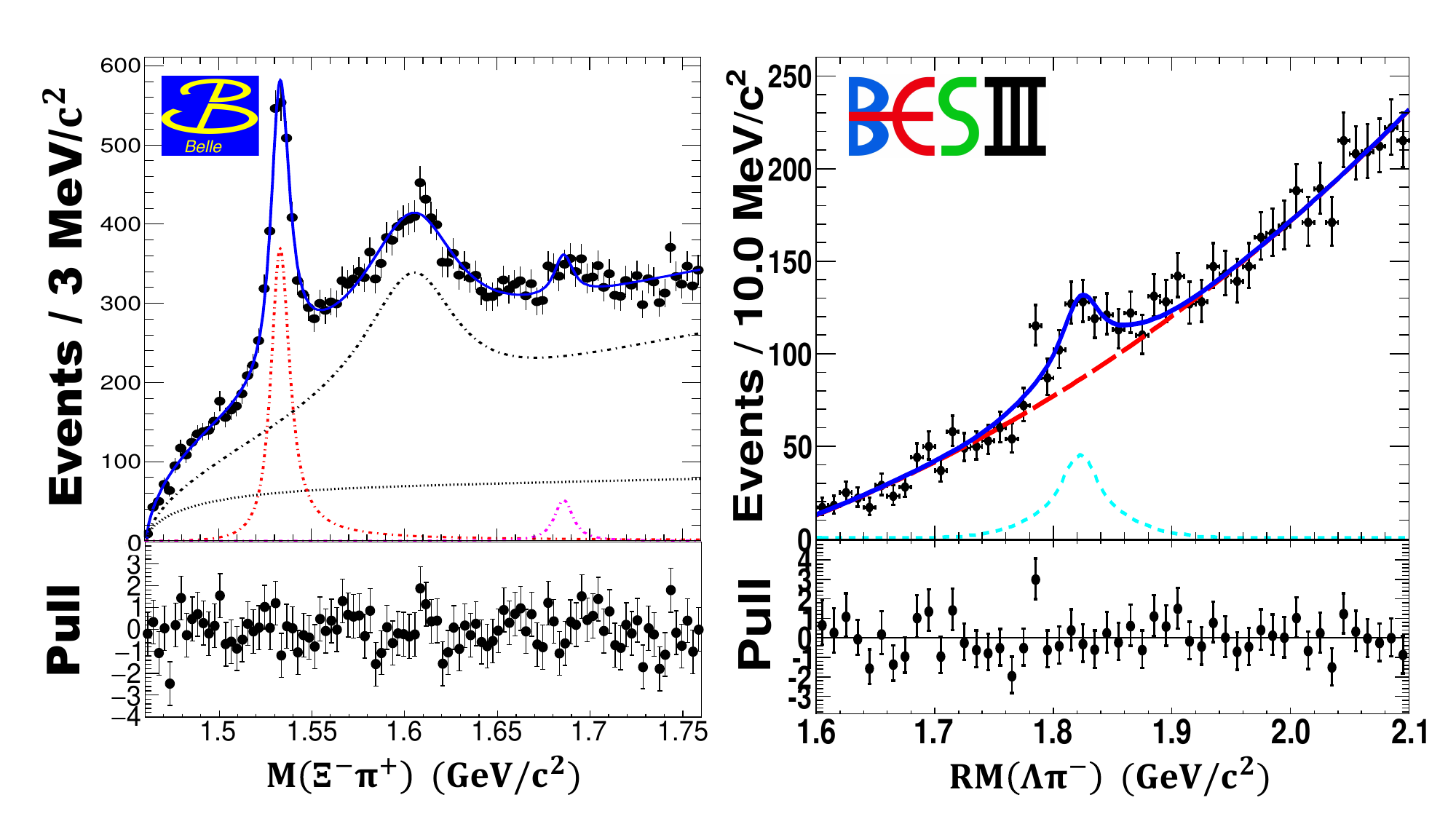}
	\end{center}
	\caption{Distributions of $M(\Xi^-\pi^+)$ from Belle experiment~\cite{Belle:2018lws} and $RM(\Lambda\pi^+)$ from BESIII experiment~\cite{BESIII:2019cuv}.}
	\label{Fig:Xi}
\end{figure}

\noindent channels such as $\Lambda K$ and $\Sigma K$. Taking $\Xi(1620)$ as an example, this particle was first discovered in the 1970s~\cite{Ross:1972bf, Bellefon:1975pvi, Briefel:1977bp}. Recently, the Belle collaboration updated its parameters by analyzing the $\Xi_c \to \Xi \pi \pi$ decay process~\cite{Belle:2018lws}, measuring a mass and width of $(1610.4 \pm 6.0 ^{+6.1}_{-4.2})~\mathrm{MeV}/c^2$ and $(59.9 \pm 4.8^{+2.8}_{-7.1})~\mathrm{MeV}$, respectively. These results have inspired theoretical investigations into the internal structure of the $\Xi(1620)$, such as the hadronic molecule picture~\cite{Huang:2021ahp}. Figure~\ref{Fig:Xi} shows the invariant mass of $\Xi^-\pi^+$. A clear $\Xi(1620)$ signal can be observed  around $1.6~\mathrm{GeV}$ with a significance greater than $25\sigma$, while only an evidence of the $\Xi(1690)$ is found to be a significance of $4\sigma$. However, these results exhibit some discrepancies with earlier experiments and have not yet been independently verified by other experiments. Currently, the BESIII collaboration is conducting further tests on the existence of $\Xi(1620)$ using the process $e^+ e^- \to \Xi^* \bar{\Xi}$. 

For higher-mass $\Xi$ excited states, the BESIII collaboration confirmed the existence of $\Xi(1820)$ in 2020 through the recoil invariant mass spectrum of $\Xi$ baryons while measuring the $e^+ e^- \to \Xi^- \bar{\Xi}^+$ cross section at $\sqrt{s}= 4.0$-$4.6~\mathrm{GeV}$ energy region~\cite{BESIII:2019cuv}. A Breit-Wigner fit yielded a mass and width of $(1825.5 \pm 4.7 \pm 4.7)~\mathrm{MeV}/c^2$ and $(17.0 \pm 15.0 \pm 7.9)~\mathrm{MeV}$, respectively. This marks the first update of the particle's parameters since the WA89 experiment~\cite{WA89:1999nsc}, with significantly improved measurement precision, though its spin-parity quantum numbers remain undetermined. 

\begin{figure}[H]
	\begin{center}
	\includegraphics[width=1.0\columnwidth]{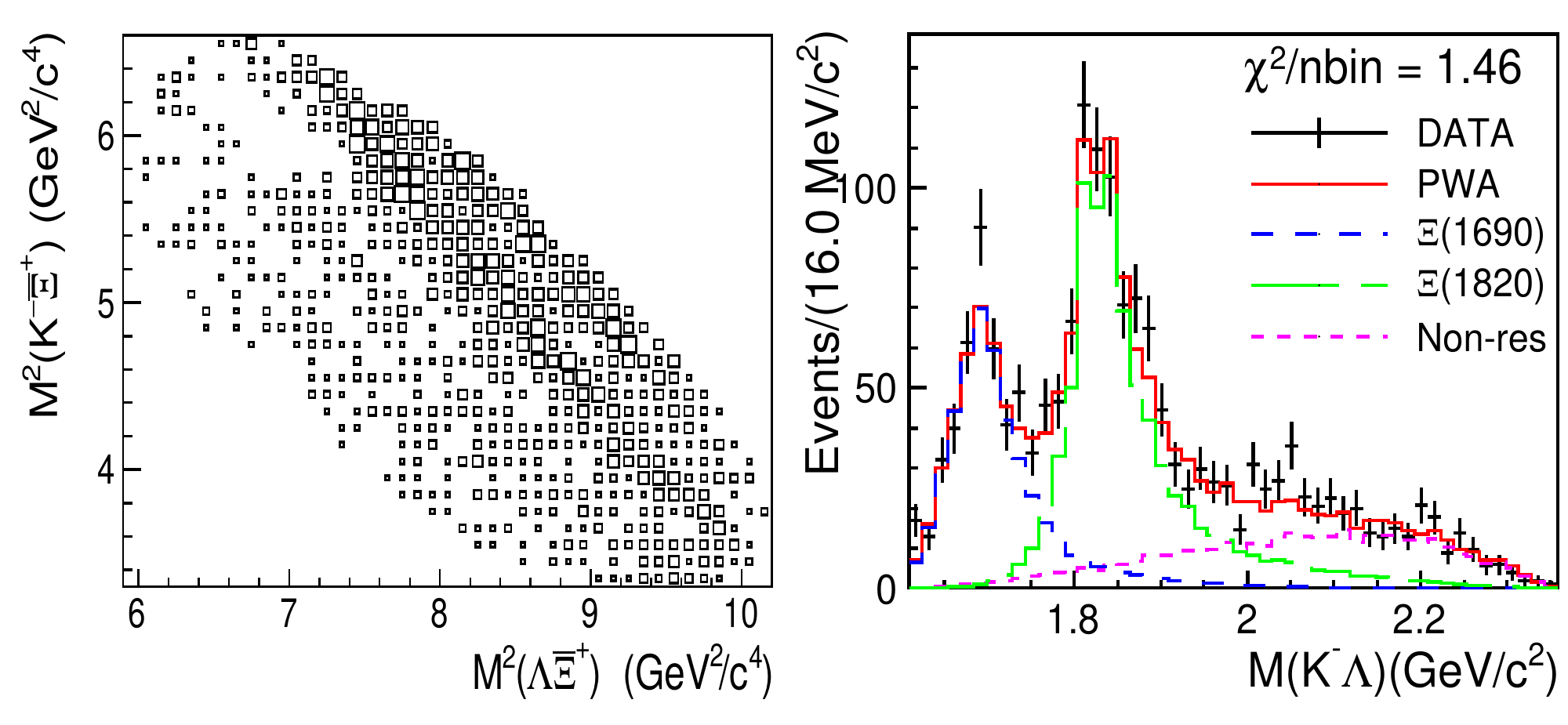}
	\end{center}
	\caption{Dalitz plot of $M^2(K^-\bar{\Xi}^+)$ versus $M^2(\Lambda\bar{\Xi}^+)$ and distribution of $M(K^-\Lambda)$ from BESIII experiment~\cite{BESIII:2023mlv}.}
	\label{Fig:Xi-02}
\end{figure}

\begin{figure}[H]
	\begin{center}
	\includegraphics[width=1.0\columnwidth]{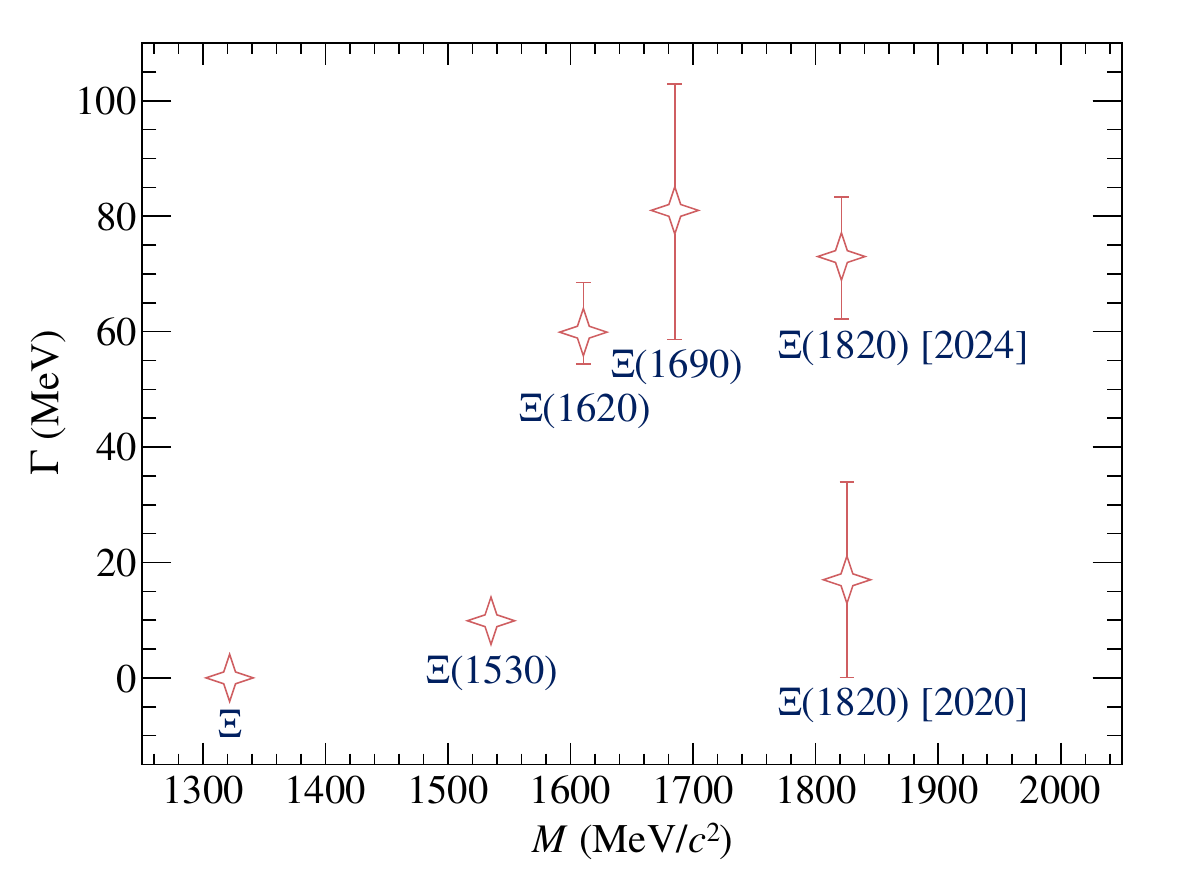}
	\end{center}
	\caption{Comparison of width and mass for $\Xi$ baryon and its excited states from the Belle and BESIII experiments.}
	\label{Fig:Xi-03}
\end{figure}

\noindent As of 2024, both excited states, $\Xi(1690)$ and $\Xi(1820)$, have been experimentally confirmed. In 2024, the $\Xi(1690)$ and $\Xi(1820)$ states were analyzed in the decay process $\psi(3686) \to \Lambda K^- \bar{\Xi}^+$ using a PWA approach~\cite{BESIII:2023mlv}. The $\Xi(1690)$ and $\Xi(1820)$ were observed with the significance of greater than 10$\sigma$ as shown in Fig.~\ref{Fig:Xi-02}. Their masses and widths were determined as $(1685^{+3}_{-2} \pm 12)~\mathrm{MeV}/c^2$ and $(81^{+10}_{-9} \pm 20)~\mathrm{MeV}$ for $\Xi(1690)$, and $(1821^{+2}_{-3} \pm 3)~\mathrm{MeV}/c^2$ and $(73^{+6}_{-5} \pm 9)~\mathrm{MeV}$ for $\Xi(1820)$, respectively. The analysis also confirmed the quantum numbers $1/2^-$ for $\Xi(1690)$ and $3/2^-$ for $\Xi(1820)$, in agreement with previous results cited by the PDG. Notably, the measured width of $\Xi(1820)$ is significantly larger than that reported in earlier work, which has prompted theoretical interest~\cite{Molina:2023uko} regarding the internal structure of this resonance. Figure~\ref{Fig:Xi-03} shows the comparison of resonance parameters for $\Xi$ excited baryon states.

\indent \emph{4.5~$\Omega^-$ excited states}. The $\Omega^-$ baryon is a hyperon composed of three strange quarks ($sss$), and the study of its excited states is an important topic in baryon 

\begin{figure}[H]
	\begin{center}
	\includegraphics[width=1.0\columnwidth]{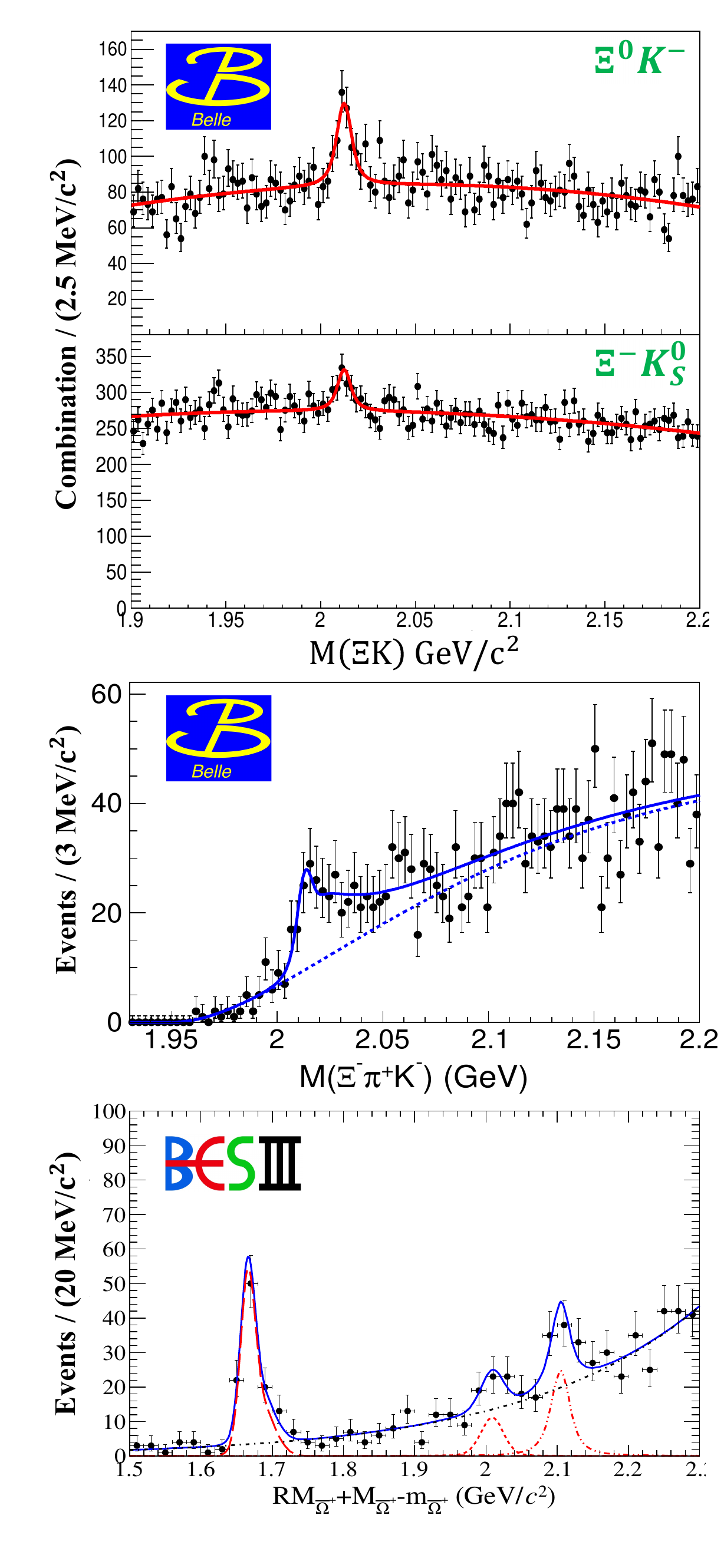}
	\end{center}
	\caption{Distributions of $M(\Xi^0K^-)$, $M(\Xi^-K_S^0)$ and $M(\Xi^-\pi^+K^-)$ from Belle experiment~\cite{Belle:2018mqs,Belle:2022mrg} and $\mathrm{RM}_{\bar{\Omega}^+} + M_{\bar{\Omega}} - m_{\bar{\Omega}}$ from BESIII experiment~\cite{BESIII:2024eqk}.
 }
	\label{Fig:Omega}
\end{figure}


\noindent spectroscopy. According to the quark model, the $\Omega^-$ baryon should possess a rich spectrum of excited states. However, experimental knowledge of these states remains extremely limited. The significant gap between theoretical predictions and experimental observations constitutes one of the core issues of the long-standing ``missing baryon resonances'' puzzle that has perplexed the particle physics community for decades. In 2018, the Belle experiment reported the first observation of an excited $\Omega^-$ baryon, i.e., $\Omega(2012)^-$, in the decay modes of $\Xi^0K^-$ and $\Xi^-K^0_S$ using the data taken in $\Upsilon(1S)$, $\Upsilon(2S)$, and $\Upsilon(3S)$ decays with a significance greater than 7$\sigma$~\cite{Belle:2018mqs}. The mass and width of $\Omega(2012)^-$ were measured to be $(2012.4\pm0.7\pm\ 0.6)~\mathrm{MeV}/c^2$ and $\Gamma=(6.4^{+2.5}_{-2.0} \pm1.6)~\mathrm{MeV}$. In 2025, Belle also observed the $\Omega(2012)^-$ baryon with $\Xi(1530)\bar{K}\to\Xi\pi\bar{K}$ with a significance of 5.2$\sigma$~\cite{Belle:2022mrg}. Figure~\ref{Fig:Omega} shows the invariant masses of $\Xi^0K^-$, $\Xi^-K_S^0$ and $\Xi^-\pi^+K^-$. The mass was measured to be $(2012.5\pm0.7\pm0.5)~\mathrm{MeV}/c^2$, in agreement with the previous result~\cite{Belle:2018mqs}. The effective couplings for $\Omega(2012)^-\to\Xi(1530)\bar{K}$ and $\Xi\bar{K}$ were found to be $(39 \pm 31 \pm 9) \times 10^{-2}$ and $(1.7 \pm 0.3 \pm 0.3) \times 10^{-2}$, respectively. Meanwhile, the Belle measurement on the narrow width implies that the quantum number of $\Omega(2012)^-$ is more favorable to $3/2^{-}$ according to the prediction of theoretical models~\cite{Capstick:1986ter,Faustov:2015eba,Loring:2001ky,Pervin:2007wa,Oh:2007cr,Engel:2013ig,Wang:2024ozz,Shen:2025xcq}.

In recent years, the BESIII experiment has made breakthrough progress in this field. Using a data sample of approximately 19 fb$^-1$ collected at c.m. energies ranging from $4.13$ to $4.70~\mathrm{GeV}$, the BESIII collaboration developed new research methods and achieved two key discoveries in the study of $\Omega^-$  baryon excited states. First, in the process $e^+e^-\to\Omega(2012)^-\bar{\Omega}^{+}$, the research team confirmed the existence of $\Omega(2012)^-$ with a significance of 3.5 standard deviations~\cite{BESIII:2024eqk}, validating the resonance state initially reported by the Belle experiment. More importantly, BESIII has, for the first time, found evidence for a new excited state of the $\Omega^-$ hyperon, with a signal significance of 4.1 standard deviations including systematic uncertainties and the look-elsewhere effect. This new particle has been designated as $\Omega(2109)^-$. Figure~\ref{Fig:Omega} (Bottom) shows the recoil mass spectrum of $\Omega^-$. The experimental measurements yield a mass of $M = (2108.5 \pm 5.2 \pm 0.9)~\mathrm{MeV}/c^2$ and a width of $\Gamma = (18.3 \pm 16.4 \pm 5.7)~\mathrm{MeV}$. These two discoveries hold significant theoretical implications. The masses of $\Omega(2012)^-$ and $\Omega(2109)^-$ are in excellent agreement with predictions from the lattice QCD calculation~\cite{Edwards:2012fx,Xie:2024wbd}, which had predicted masses of approximately $2.0~\mathrm{GeV}/c^2$ and $2.1~\mathrm{GeV}/c^2$ for $\Omega^-$ excited states with spin-parity $J^{P} = 1/2^{-}$ and $3/2^{-}$, respectively. Meanwhile, the experimentally measured mass of $\Omega(2109)^-$  exhibits a deviation of approximately $50~\mathrm{MeV}$ below the theoretical predictions~\cite{Luo:2025cqs}. 

\noindent In addition, the theoretical study of Hamiltonian effective field theory suggests that the quantum number of $\Omega(2012)^-$ could be $J^P=3/2^-$ resonance while recently reported $\Omega(2109)^-$ may be $J^P=1/2^-$~\cite{Han:2025gkp}. The experimental achievements not only provide crucial experimental evidence for understanding the internal structure of $\Omega^-$ baryon excited states, but also lay an important foundation for ultimately resolving the ``missing baryon resonances'' puzzle and refining 

\begin{figure}[H]
	\begin{center}
	\includegraphics[width=1.0\columnwidth]{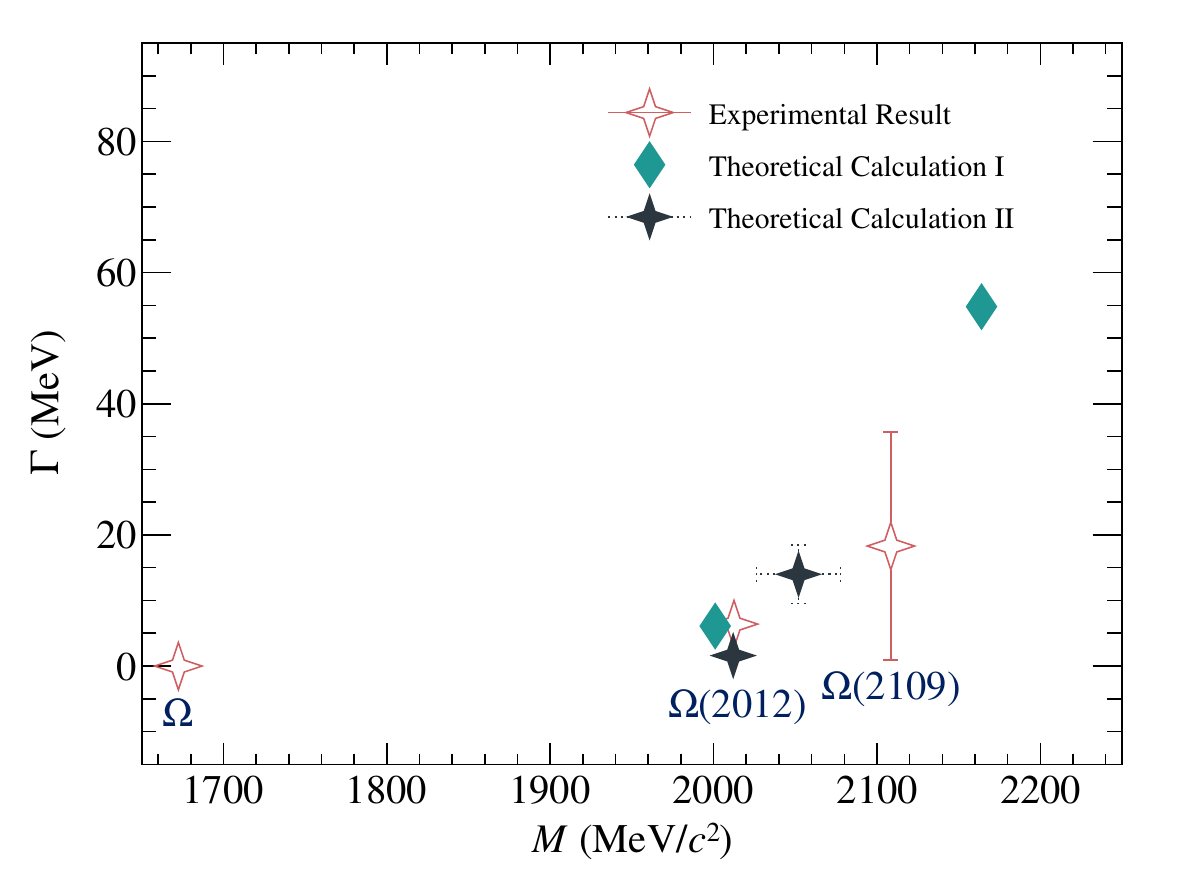}
	\end{center}
	\caption{Comparison of resonance parameters for $\Omega^-$ baryon and its excited states from the Belle and BESIII experiments, where the theoretical calculations are cited from Ref.~\cite{Luo:2025cqs} and Ref.~\cite{Han:2025gkp} for the results I and II, respectively.}
	\label{Fig:Omega-02}
\end{figure}

\noindent QCD theory. In the future, with more data accumulation and in-depth analysis, further breakthroughs in $\Omega^-$ baryon spectroscopy are anticipated. Figure~\ref{Fig:Omega-02} shows the comparison of resonance parameters for $\Omega^-$ excited baryon states between the experimental measurements and theoretical predictions.

\indent \emph{5.~Summary}. Over the past decade of operation, the BESIII experiment has achieved remarkable accomplishments in the field of baryon spectroscopy. The work of the BESIII collaboration is systematically deepening our understanding of baryon structure. In the study of light-flavored baryon excited states, BESIII has made a series of breakthroughs in the investigation of $\Sigma$, $\Xi$, and $\Omega^-$ hyperon excitations by utilizing 10 billion $J/\psi$ and 3 billion $\psi(3686)$ events~\cite{BESIII:2020nme}. These results provide crucial experimental data for addressing the ``missing baryon resonances'' problem, showing strong consistency with lattice QCD predictions and robustly supporting the quark model's description of the baryon spectrum. At the same time, they reveal spin-orbit coupling effects that have not been sufficiently considered in the quark model.

Looking ahead, the BESIII experiment continues to accumulate data, with its c.m. energy coverage continuously expanding. According to BESIII's future physics plan, high-precision measurements will continue in the tau-charm energy region during the remaining operation period, with optimized data acquisition strategies and physics motivations for potential luminosity upgrades. Particularly promising is the Super Tau-Charm Factory in Russia~\cite{Levichev:2018cvd} and China~\cite{Achasov:2023gey}, which is at the forefront of higher luminosity and has entered the preliminary research stage. Its designed luminosity will be two orders of magnitude higher than that of BEPCII, and it is expected to accumulate a data sample two orders of magnitude larger than that of BESIII. This will advance baryon spectroscopy research to unprecedented precision, making the discovery of more excited states possible. Complementary to these high-energy spectroscopy studies, future facilities such as HIAF in China will also drive significant progress in hypernuclear physics, providing unique insights into hyperon-nucleon interactions and few-body systems containing strange quarks~\cite{Chen:2025eeb}. Solving the mystery of the ``missing baryon resonances'', refining non-perturbative QCD theory, and testing the precision of the Standard Model. These core objectives of baryon physics are gradually being approached with each new discovery from the BESIII experiment. From $\Sigma$ hyperon excitations to the $\Omega^-$ hyperon spectrum, the BESIII experiment is writing a new chapter in the study of baryon spectroscopy.
\indent \emph{Acknowledgements}.
This work was supported by 
the Fundamental Research Funds for the Central Universities Nos.
lzujbky-2025-ytA05, lzujbky-2025-it06, lzujbky-2024-jdzx06;
the Natural Science Foundation of Gansu Province No. 22JR5RA389, No.25JRRA799;
the `111 Center' under Grant No. B20063;
the National Natural Science Foundation of China under Contract No. 12225509,12247101.

\bibliography{ref}
\bibliographystyle{cplrev2}

\end{multicols}
\end{document}